\documentclass[reprint,amsmath,amssymb,aps,floatfix,prl,superscriptaddress]{revtex4-1}

\usepackage{graphicx}
\usepackage{dcolumn}
\usepackage{bm}
\usepackage{epstopdf}
\usepackage{pifont}
 
\begin{document}

\preprint{}

\title{Slow plastic creep of 2D dusty plasma solids}

\author{Peter Hartmann}
\affiliation{Institute for Solid State Physics and Optics, Wigner Research Centre, Hungarian Academy of Sciences, P.O.B. 49, H-1525 Budapest, Hungary}
\affiliation{Center for Astrophysics, Space Physics and Engineering Research (CASPER), One Bear Place 97310, Baylor University, Waco, TX 76798, USA}

\author{Anik\'o Zs. Kov\'acs}
\affiliation{Institute for Solid State Physics and Optics, Wigner Research Centre, Hungarian Academy of Sciences, P.O.B. 49, H-1525 Budapest, Hungary}

\author{Angela M. Douglass}
\affiliation{Ouachita Baptist University, 410 Ouachita St., Arkadelphia, AR 71923, USA}
\author{Jorge C. Reyes}
\author{Lorin~S.~Matthews}
\author{Truell W. Hyde}
\affiliation{Center for Astrophysics, Space Physics and Engineering Research (CASPER), One Bear Place 97310, Baylor University, Waco, TX 76798, USA}

\date{\today}

\begin{abstract}
We report complex plasma experiments, assisted by numerical simulations, providing an alternative qualitative link between the macroscopic response of polycrystalline solid matter to small shearing forces and the possible underlying microscopic processes. In the stationary creep regime we have determined the exponents of the shear rate dependence of the shear stress and defect density, being $\alpha = 1.15 \pm 0.1$ and $\beta = 2.4 \pm 0.4$, respectively. We show that the formation and rapid glide motion of dislocation pairs in the lattice are dominant processes. 
\end{abstract}


\maketitle

The direct in situ observation of dynamical processes in bulk condensed matter is not yet available with atomic resolution in both space and time. Femtosecond pump-probe techniques can resolve atomic motion, atomic force microscopy can detect atoms on surfaces, diffraction methods provide information about the bulk structure, but as long no method can combine all the benefits of these techniques, we are limited to rely of phenomenological models and numerical simulations. Alternative experimental methods have already proven to be helpful for the qualitative understanding of classical collective phenomena. Charged colloids suspended in a liquid environment, and dusty plasmas (solid micron sized particles charged and levitated in gas discharge plasmas) are both interacting many-particle systems that show very similar properties to conventional atomic matter, but at time and distance scales easily and directly accessible with simple video microscopy techniques. Both methods provide insight into the microscopic (particle level) details of different phenomena. Colloid systems are characterized by over-damped dynamics, due to the liquid environment, which makes them well suited for structural and phase transition studies \cite{NatPhys}, while the weak damping in low pressure gas discharges makes dusty plasmas perfect for studies of wave-dynamics, instabilities, and other collective excitations \cite{RotoDust}. 

In material science and metallurgy, creep is the time dependent plastic strain at constant stress and temperature; therefore, it is a special type of plastic deformation of solid matter. In general it is a slow process driven by the thermally activated movement of dislocations (dislocation creep), vacancies (vacancy creep) or diffusion (Nabarro-Herring and Coble creep). The applied stresses are below the rapid yield stress resulting in atomic movements that are crystallographically organized. The applied temperatures are usually above $\frac12 T_\text{m}$, where $T_\text{m}$ is the melting temperature. The time ($t$) evolution of the deformation (strain $\varepsilon$) at constant stress is often described by one of the empirical formulae 
\begin{eqnarray}
\varepsilon &=& \varepsilon_0 + \delta \ln t + \phi t ~~~\text{or} \label{eq:trans1}\\
\varepsilon &=& \varepsilon_0 + \vartheta t^{1/3} + \phi t, \nonumber
\end{eqnarray}
where $\varepsilon_0$ is the immediate strain and $\delta$, $\vartheta$ and $\phi$ are creep coefficients \cite{Cottrell}. After a short transient phase (``primary creep''), approximated as logarithmic ($\sim \delta \ln t$) or using Andrade's law ($\sim \vartheta t^{1/3}$), this describes a steady-state ``secondary'' creep, dominated by the last term, where the rate $\phi$ is determined by the balance of work hardening and thermal softening. Under such circumstances the steady state creep is fairly well represented by the Norton equation: 
\begin{equation}\label{eq:Weert}
\dot\varepsilon = \frac{\partial \varepsilon}{\partial t} = C(T)\sigma^\alpha,
\end{equation}
where $\sigma$ is the stress in the system, $C(T)$ is a factor characteristic for the material and the experimental conditions (temperature, grain size of a polycrystalline sample, Young modulus, etc.) and $\alpha$ is the Norton exponent with $1\leq \alpha \leq 10$ (from experiment, depending on the dominating microscopic process). Simplified theoretical models reduced to thermal activation of independent dislocation glide movements in the lattice (Harper-Dorn creep) predict an exponent $\alpha=1$ in eq. (\ref{eq:Weert}) \cite{Cottrell,Harper58,Dorn1,Kassner07}. This phenomenological description was derived mostly for tension strain, but can be applied for shear (along the $x$-axis in our notation) as well, with the substitution $\varepsilon \rightarrow \gamma$, where $\gamma = \partial x/\partial y$, and $\dot\gamma = \partial v_x/\partial y$. In the case of large grained pure materials (mostly metals) at intermediate stresses and temperatures, the deformation is realized by the creation and glide movement of edge dislocations with Burgers vectors $\vec{b}$, resulting in shear rates $\dot\gamma = \rho\vec{b}\vec{v}$ (Orowan equation), where $\rho$ is the density of mobile dislocations with average velocity $\vec{v}$. In this model the dislocation density $\rho$ is expected to depend on the shear stress as 
\begin{equation}\label{eq:beta}
\rho \propto \sigma^\beta.
\end{equation}
Simple theoretical arguments and tensile experiments on single and polycrystalline copper predict an exponent $\beta \approx 2$ \cite{Courtney}.

The effect of shearing forces and liquid state viscosity were investigated in early years of dusty plasma research \cite{Ivisc,Nosvisc,Gavvisc,Donko06}. More recently, effects of shearing forces on the microstructure were experimentally investigated in greater detail including liquid flows and plastic deformations of a crystalline solid \cite{Idet1,Idet,Samsdet,Nosdet,HartiPRE11,Nosenko13}. Here we present dusty plasma experiments providing the link between the standard, macroscopic measures used in material sciences and metallurgy to describe plastic deformations, and the detailed microscopic information provided by dusty plasma experiments.

\begin{figure}[htb]
\includegraphics[width=1\columnwidth]{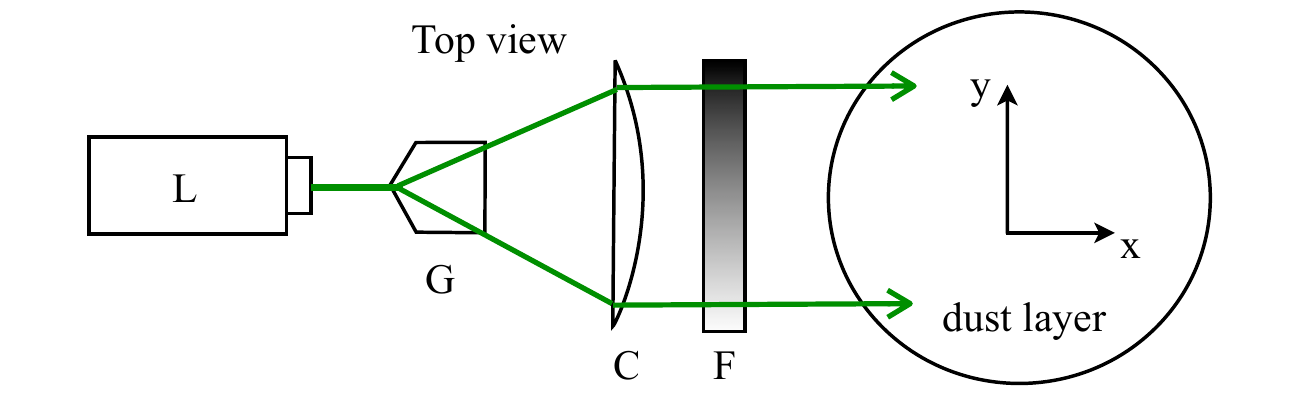}
\caption{\label{fig:setup} 
(color online) Scheme of the optical setup used to generate the wide shearing laser beams. Main parts: L -- Laser source (TEX: max 5.5W @ 532nm; BUD: max. 1W @ 440nm), G -- laser line generator lens, C -- cylindric lens, F -- linearly variable density filter.}
\end{figure}

We have carried out two independent experiments on single layer crystalline dusty plasma systems aiming to investigate the microscopic details of the shearing creep deformation. The first experiment was performed in the Hypervelocity Impacts and Dusty Plasmas Lab (HIDPL) of the Center for Astrophysics, Space Physics, and Engineering Research (CASPER) at Baylor University, Waco, Texas, while the second series was carried out at the Institute for Solid State Physics and Optics, part of the Wigner Research Centre for Physics, Budapest, Hungary (referenced as ``TEX'' and ``BUD'' in the following). Details of the dusty plasma apparatuses and data processing techniques can be found in earlier publications \cite{Boess04,HartiPRE11}. In both cases the dust layer was illuminated by an extended, spatially linearly modulated laser sheath introducing an external force $F_x(y) = F_0(y-y_0)$ on every particle, as illustrated in figure~\ref{fig:setup}. 

Several video sequences with different powers of the manipulating laser $P_L$ were recorded, including the $P_L=0$ case, which was used to extract the wave spectra of thermally excited density fluctuations, which fitted with lattice phonon dispersion curves of 2D Yukawa systems \cite{Peeters87,IEEEYuk} resulted in system parameters used during the evaluation of the $P_L > 0$ data. These parameters are collected in table I. 

\begin{table}[htb]
\label{table:system} 
\setlength{\tabcolsep}{11pt}
\begin{tabular*}{1\columnwidth}{ l | c c }
\toprule
parameter & TEX & BUD \\
\colrule
 $p$, Ar gas pressure [Pa] & 9.6 &  1.25 \\
 $P_\text{RF}$, RF power at 13.56 MHz [W] & 3.5 & 15 \\
 $d$, MF particle diameter [$\mu$m] & 6.37  & 9.16 \\
 $m$, particle mass [$10^{-13}$kg] & 2.04 & 6.08 \\
 $N$, particle number in field of view  & 1404  & 2810 \\
 frames per second (fps) & 60  & 29.7 \\
 recorded frames per experiment & 5500 & 17000 \\
 resolution [$\mu$m/pixel] & 35.2 & 30 \\
 $q$, charge [$e$, elementary charge] & 7250 & 13600 \\
 $n$, density [$10^6$m$^{-2}$] & 2.76 & 2.99 \\
 $a$, Wigner-Seitz radius [$10^{-4}$m] & 3.4 & 3.26 \\
 $\omega_p = q\sqrt{n/2\epsilon_0 m a}$ [rad/s] & 55.1 & 63.5 \\
 $\lambda_\text{D}$, Debye screening length [$10^{-4}$m] & 4.9 & 4.8 \\
 $\kappa = a/\lambda_\text{D}$, Yukawa parameter  & 0.7 & 0.68 \\
 $T/T_\text{melt}$ from $g(r)$, see \cite{OttGamma} & 0.42 &0.37 \\ 
 $v_\text{th}$, avg. thermal speed [$10^{-4}$m/s] & 12.7 & 4.46 \\
 $\dot\gamma_0 = v_\text{th}/a$ [s$^{-1}$] & 3.74 & 1.37 \\
 $\sigma_0 = m n \omega_0 a v_\text{th}$ [$10^{-11}$kg/s$^2$] & 1.34 & 1.68 \\
\botrule
\end{tabular*}
\caption{System parameters derived from the structural and dynamical properties of the unperturbed systems. Uncertainties are estimated to vary from 1\% (eg. $m$, $d$, resolution) up to 10\% (eg. $q$, $\kappa$)}
\end{table}

\begin{figure}[t!]
\includegraphics[width=1\columnwidth]{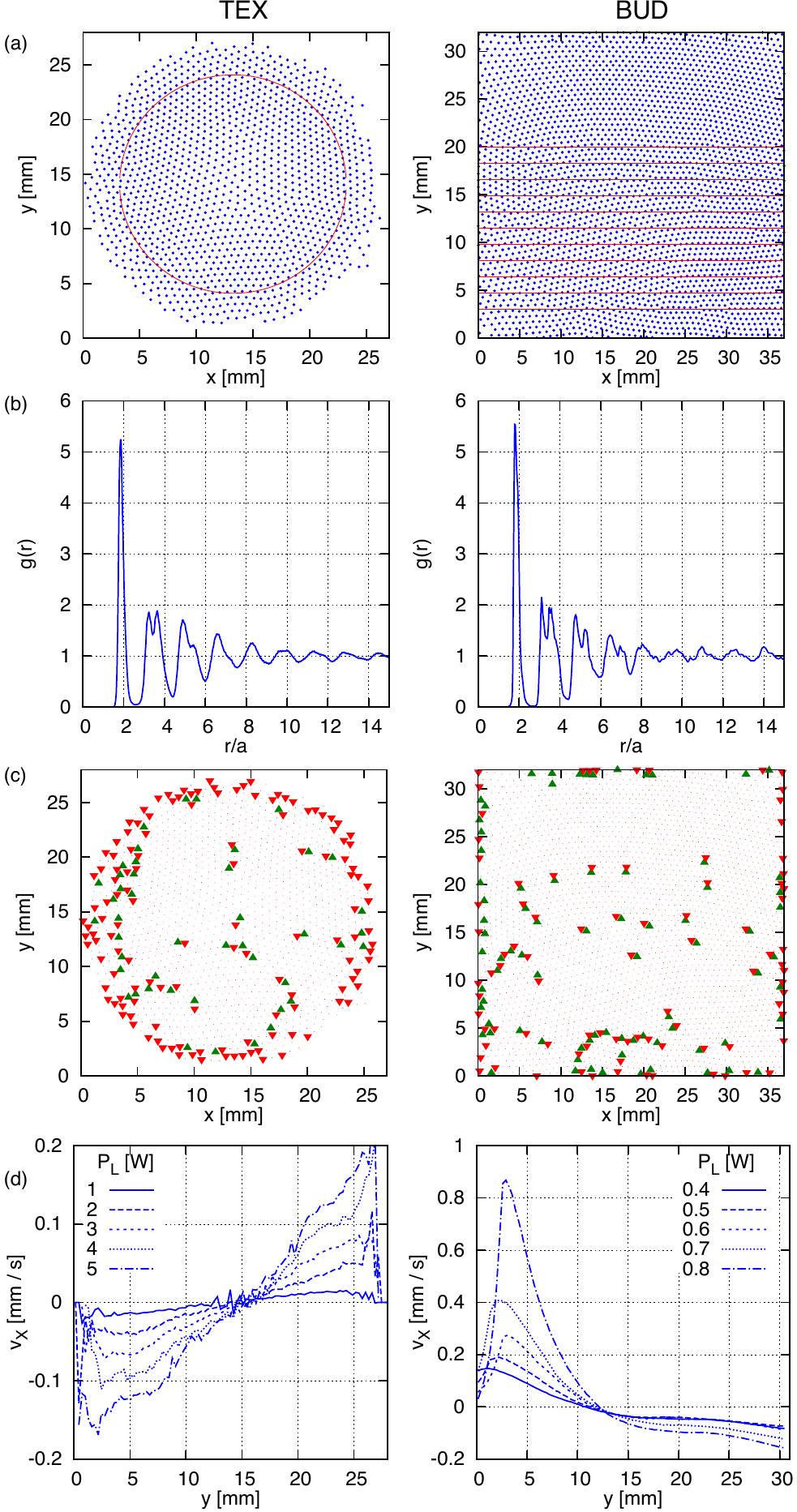}
\caption{\label{fig:eval} 
(color online) Illustration of the data evaluation process for both experiments (left column: TEX, right column: BUD). For the unperturbed cases: (a) example of particle snapshots with the regions of interest; (b) pair correlation function; (c) defect maps (\ding{115}: particle with 7 neighbors; \ding{116}: particle with 5 neighbors). (d) $v_x(y)$ transverse particle velocity profiles for different shearing laser powers.}
\end{figure}

The elementary steps of the data evaluation process are illustrated in figure~\ref{fig:eval} and include (a) the identification of the particles and determination of the regions of interest, (b) measurement of the pair distribution function $g(r)$, (c) performing Delaunay triangulation to determine true nearest neighbor bonds and identify defects, defined as particles with neighbor numbers $\ne 6$. The measurement of the particle velocities includes the determination of flow, peculiar and particular thermal velocities. Examples of the induced flow velocity profiles $v_x(y)$ for laser intensities in the range of $0.4\dots5$~W are presented in figure~\ref{fig:eval}(d), showing a pronounced difference between the two experiments. Linear velocity profiles were found in the TEX case, while in the BUD experiments the velocity profiles decayed exponentially. To benefit from both situations we have defined the regions of interest (marked with red lines in figure~\ref{fig:eval}(a)). TEX: to avoid complication with the boundaries (including circulations) of the dust clouds during the subsequent evaluation steps we consider only particles within a circle with a diameter of about 75\% of that of the cloud. The shear rate $\dot\gamma = \partial v_x/\partial y$ is constant over the whole $y$ range. BUD: as the dust cloud is much larger than the field of view of the camera, boundary effects are of minor concern. On the other hand, the velocity profiles show strong non-linear shape. We define 10 slabs within the decaying tail (see figure~\ref{fig:eval}(a) right column) of $v_x(y)$, where the shear rate within each slab can be approximated as constant, but it changes from slab to slab, providing the possibility of evaluating 10 different $\dot\gamma$ values at once. The qualitative difference of the velocity profiles is assumed to be related to the significantly different Ar gas pressures, which, through friction is responsible for the energy dissipation.

The two quantities of interest, besides the directly measured shear rate, are the defect fraction (proportional to the dislocation density) and the shear stress in the system. The former results from the Delaunay triangulation. The latter is computed applying the formula
\begin{equation}\label{eq:P}
\sigma_{xy} = \frac{1}{A}\bigg[ \sum_i m v_{i,x} v_{i,y} + \frac{1}{2}\sum_i\sum_{j\ne i} r_{ij,y} F_{ij,x}\bigg],
\end{equation}
where $A$ is the area of the region of interest, $v_{i,x}$ is the $x$ component of the peculiar velocity of particle $i$, $r_{ij,y}$ is the $y$ component of the distance vector between particles $i$ and $j$, and $F_{ij,x}$ is the $x$ component of the force acting on particle $i$ due to pair interaction with particle $j$. Summation over $i$ is for particles within $A$, while summation of $j$ runs over all particles interacting with particle $i$. To approximate the inter-particle forces, we use the widely accepted Yukawa (screened Coulomb) model, with a pair-potential: $\Phi(r) = q \exp[-r/\lambda_D]/(4 \pi\varepsilon_0 r)$, where $q$ is the charge of the dust particle, $\lambda_D$ is the Debye screening length, and $\varepsilon_0$ is vacuum permittivity. The values of these quantities are listed in Table~I.

\begin{figure}[htb]
\includegraphics[width=1\columnwidth]{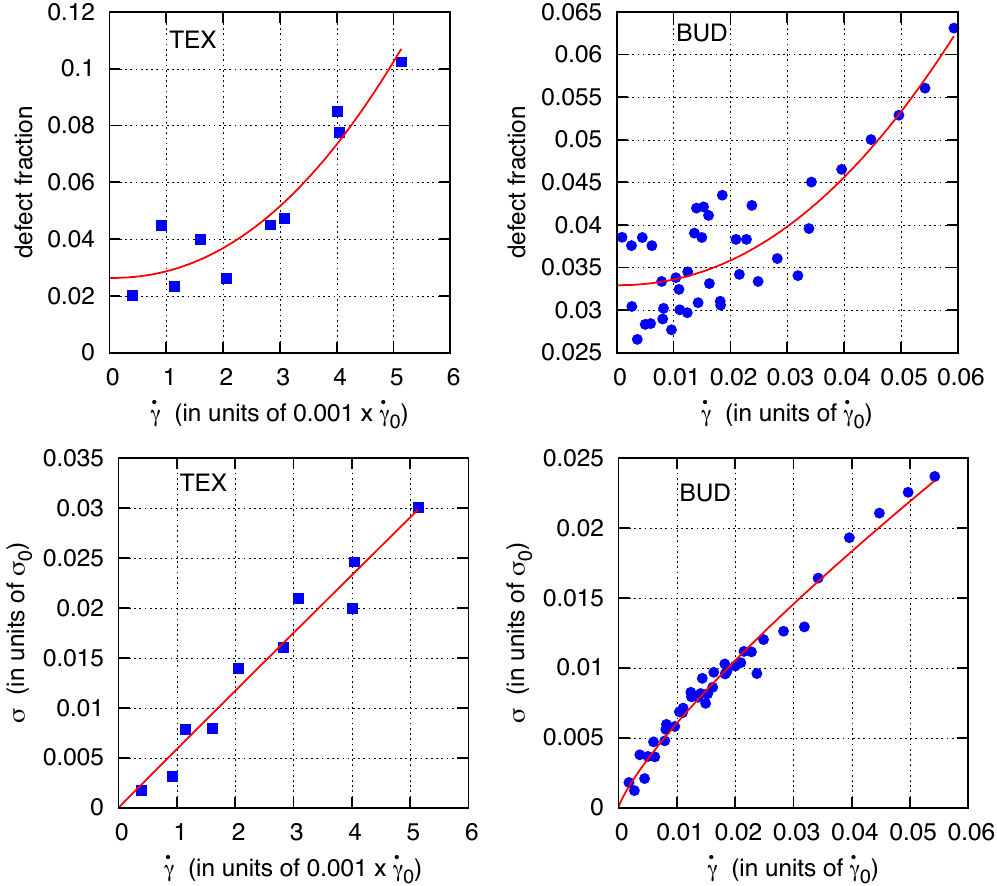}
\caption{\label{fig:gamma} 
(color online) Defect fraction (upper row) and shear stress (lower row) versus shear rate from both the TEX (left column) and BUD (right column) experiments. Lines are functional fits, see text.}
\end{figure}

The results of the data evaluation are presented in figure~\ref{fig:gamma}. The experimental data points are approximated by functions having the forms: 
\begin{equation}
D\left(\dot\gamma\right) = D_0 + d \dot\gamma^b, ~~\text{and}~~~
\sigma \left(\dot\gamma\right) = c \dot\gamma^f. 
\end{equation}
In a polycrystalline systems at finite temperature, even without external shear, the equilibrium defect (dislocation) density $D_0$ is non-zero, in contrast to the shear stress. $D_0$ corresponds to the immobile fraction of the dislocations, which are excluded from eq.~\ref{eq:beta}. Least-squares fitting resulted in $b_\text{TEX}=2.15\pm 0.5$, $b_\text{BUD}=2.1\pm 0.3$, $f_\text{TEX}=0.95\pm 0.1$, $f_\text{BUD}=0.8\pm 0.03$ for the exponents. Combining these results, we find that the defect density -- shear stress relation, as given in eq.~(\ref{eq:beta}) is, at least in the studied shear rate regime, a good approximation and has an exponent $\beta = b/f \approx 2.4 \pm 0.4$. Furthermore, the Norton exponent in eq.~(\ref{eq:Weert}) is $\alpha= 1/f \approx 1.15 \pm 0.1$. These results are consistent with both (BUD and TEX) experiments. Taking into account the fact that the crystalline domains are large, compared to the system size, these values of the exponents suggest the Harper-Dorn creep \cite{Cottrell,Harper58,Dorn1} to be the dominant process. 

\begin{figure}[htb]
\includegraphics[width=1\columnwidth]{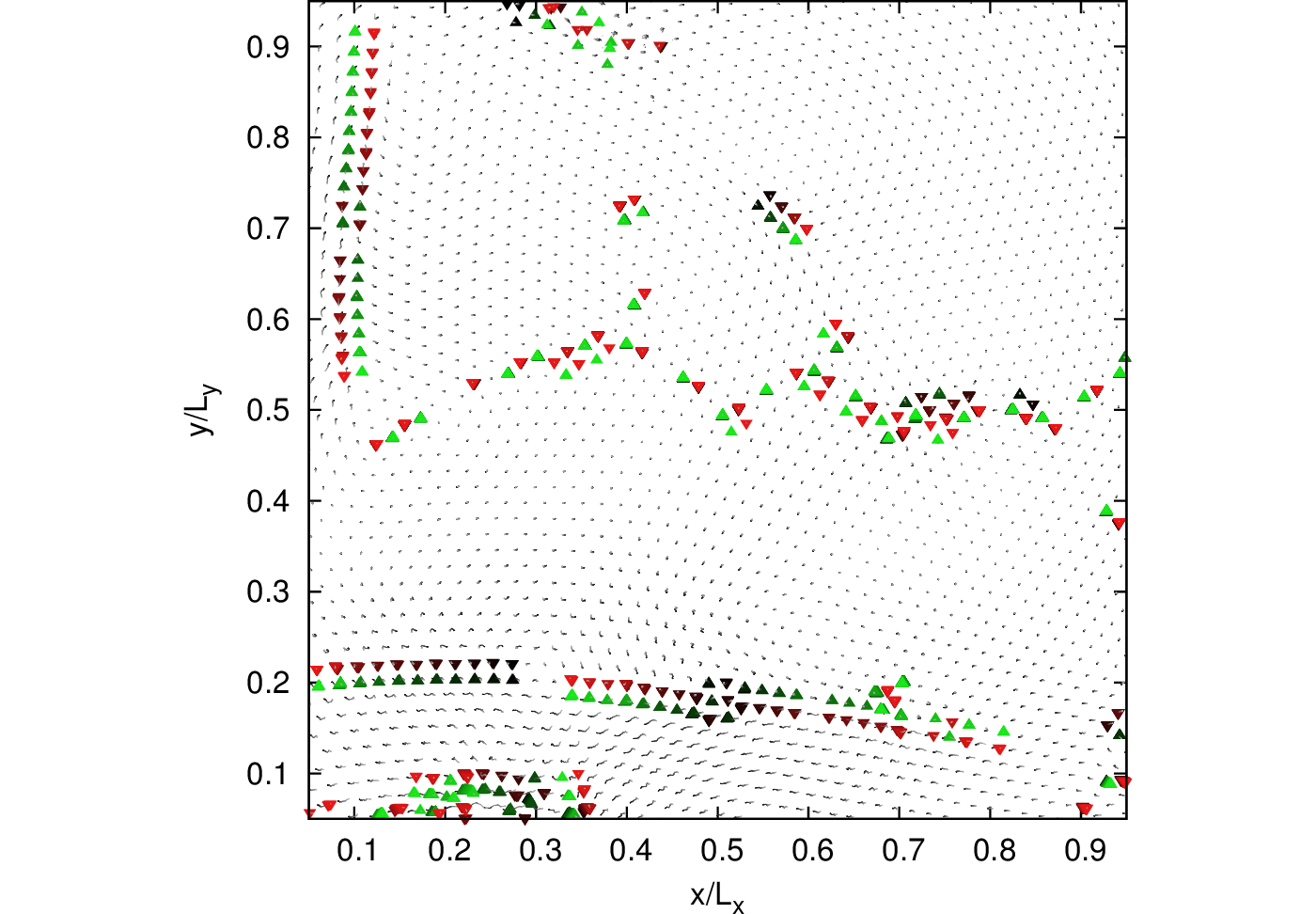}
\caption{\label{fig:disl} 
(color online) Defect maps of subsequent system snapshots from the BUD experiment. Colors lighten with elapsing time. \ding{115}: particles with 7 neighbors; \ding{116}: particles with 5 neighbors.}
\end{figure}

To support this statement, fig.~\ref{fig:disl} shows an overlay of subsequent defect maps (BUD, $P_L=0.6$~W). This illustrates the formation and rapid glide of dislocation pairs through the lattice, a typical scenario occurring frequently in the system. In frustrated dusty plasma crystals such individual processes have already been observed \cite{NosDisloc,Durniak} and studied numerically in detail \cite{Zhdanov12}.

In addition to dusty plasma experiments, non-equilibrium molecular dynamics (NEMD) simulations \cite{Rapaport} for a 2D particle ensemble with N = 46000 particles were conducted, again assuming a Yukawa inter-particle interaction potential. Particles were placed randomly in a rectangular simulation box to match the ground state hexagonal lattice. After a thermalization period, a homogeneous shear algorithm was applied to generate planar Couette flow \cite{Evans}. In the algorithm the shear flow is induced with the Lees-Edwards (sliding brick) periodic boundary conditions, which is used in conjunction with the Gaussian thermostatted SLLOD equations of motion:
\begin{equation} \label{eqofmotion}
\frac {d\mathbf{r}_i}{dt} = \frac{\tilde{\mathbf{p}}_i}{m} + {\dot\gamma}y_i\hat{\mathbf{x}},~~~~
\frac {d\tilde{\mathbf{p}}_i}{dt} = \mathbf{F}_i - {\dot\gamma}\tilde{\mathbf{p}}_{yi}\hat{x}- {\tau}\tilde{\mathbf{p}}_i, 
\end{equation} 
where $\tilde{\mathbf{p}}=(\tilde{p}_x, \tilde{p}_y)$ is the peculiar momentum of particles, $\hat{\mathbf{x}}$ is the unit vector in $x$ direction and $\tau$ is the Gaussian thermostatting multiplier, calculated in a way to ensure constant peculiar kinetic energy. The integration of this set of equations is solved by the operator-splitting technique \cite{Pan}. During the simulations we track the positions and velocities of particles and perform the analysis already introduced for the experiments.

The simulations require as dimensionless input parameters the homologous temperature $T/T_\text{m}$, and $\bar{\dot\gamma} = \dot\gamma\cdot(a/v_\text{th,m})$, where $v_\text{th,m} = \sqrt{2k_\text{B}T_m/m}$ is the thermal velocity at melting temperature. Calculations are performed for Yukawa screening parameter  $\kappa = a / \lambda_D = 2$, $T/T_\text{m} = 0.18 \dots 0.83$, and $\bar{\dot\gamma} = 0.05 - 0.5$.

\begin{figure}[htb]
\includegraphics[width=1\columnwidth]{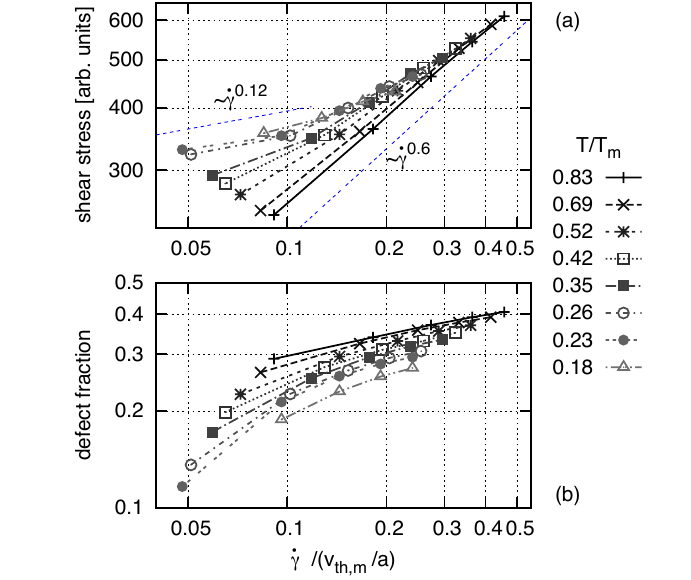}
\caption{\label{fig:MD} 
(color online) Computed shear stress (a) and defect fraction (b) versus shear rate for a set of homologous temperatures. Blue dotted lines in (a) show simple power law functions with exponents as labelled.}
\end{figure}

Figure~\ref{fig:MD} shows the computed shear stress and defect fraction values versus the strain rate for a set of homologous temperatures. The shear stress curves in fig.~\ref{fig:MD}(a) asymptotically join to a universal power-law function with a Norton exponent of $\alpha \approx 1/0.6 \approx 1.7$ at high shear rates. The effect of temperature is most dominant at low shear rates, where the system hardens with decreasing temperature, while reaching Norton exponents up to $\alpha \approx 1/0.12 \approx 8$ at the lowest investigated temperature. The defect fraction in fig.~\ref{fig:MD}(b) shows a similar, but somewhat weaker asymptotic trend. The tendency of developing fewer defects at lower temperatures in a steady-state configuration still dominates over the effect of the external frustration.

In summary, we have experimentally determined the plastic properties of 2D dusty plasma single layers in the polycrystalline phase. The macroscopic measures, like the defect density -- shear stress relation and Norton exponent during slow deformations (creep) are fully coherent with empirical models used to describe ordinary materials. This fact supports the concept of using dusty plasmas as magnified model systems of ordinary matter providing the possibility of studying a wide variety of many-body phenomena on the particle (quasi-atomic) level with spatial and temporal resolutions easily accessible on the human scale.

\begin{acknowledgments}
This research was supported by the CASPER project, and Hungarian Grant OTKA NN-103150. 
\end{acknowledgments}


%

\end{document}